\documentstyle[aps,amssymb,multicol]{revtex}
\begin{document}
\author{Z. Zhai, Patanjali V. Parimi, and S. Sridhar}
\title{Non--linear Microwave Impedance of short and long Josephson Junctions}
\address{Physics Department, Northeastern University, 360 Huntington Avenue, Boston,\\
MA 02115}
\date{\today}
\maketitle

\begin{abstract}
The non-linear dependence on applied $ac$ field ($b_{\omega }$) or current ($%
i_{\omega }$) of the microwave (ac) impedance $R_{\omega }+iX_{\omega }$ of
both short and long Josephson junctions is calculated under a variety of
excitation conditions. The dependence on the junction width is studied, for
both field symmetric (current anti-symmetric) and field anti-symmetric
(current symmetric) excitation configurations.The resistance shows step-like
features every time a fluxon (soliton) enters the junction, with a
corresponding phase slip seen in the reactance. For finite widths the
interference of fluxons leads to some interesting effects which are
described. Many of these calculated results are observed in microwave
impedance measurements on intrinsic and fabricated Josephson junctions in
the high temperature superconductors, and new effects are suggested. When a $%
dc$ field ($b_{dc}$) or current ($i_{dc}$) is applied, interesting phase
locking effects are observed in the ac impedance $Z_{\omega }$. In
particular an almost periodic dependence on the dc bias is seen similar to
that observed in microwave experiments at very low dc field bias. These
results are generic to all systems with a $\cos (\phi )$ potential in the
overdamped limit and subjected to an ac drive.
\end{abstract}

\section{INTRODUCTION}

A Josephson junction (JJ) displays several remarkable properties which are
well known \cite{Likharev,Barone}. In the context of high temperature
superconductors (HTS) the analysis of JJ has received new impetus for
several reasons. In addition to the interest in fabricating barrier
junctions made of HTS, there are new features observed in the
electromagnetic response of the HTS materials which can be attributed to the
occurrence of JJ distributed in the material. Many of the microwave
properties of ceramics are ascribed to the presence of so-called ``weak
links'' whose response is JJ-like. Josephson junction phenomena have been
observed and attributed to Josephson coupling along the c-axis between
superconducting layers\cite{Mueller}, and even within the ab-plane \cite
{Zhai97}.

Much of the analysis of an elemental JJ has been motivated by quantities
which are experimentally accessible. The best studied parameters are the 
{\em zero-frequency} or $dc$ I-V characteristics, including the Shapiro
steps seen in these $dc$ properties when the junction is irradiated by
microwaves. These analyses have been extended to study Josephson junction
arrays \cite{array} which can be fabricated artificially using low $T_{c}$
materials, and are relevant in c-axis studies.

Recently, the {\em high frequency} or microwave impedance of single crystals
has shown a remarkable non-linear dependence on the probing microwave field
which can be described in terms of Josephson-like response to purely $ab$%
-plane currents \cite{Zhai97}. Results such as these in single crystals, and
in fabricated bicrystal thin film junctions \cite{oates96}, has resulted in
the observation of new parameters, such as the {\em dynamic }or {\em ac}
(microwave) resistance and reactance. These quantities were not addressed
extensively previously because they were not of experimental interest until
recently. It is therefore timely to carry out a detailed study of the {\em %
high frequency} properties of JJ, and the results of such a study are
reported in this paper.

In this paper we present numerical simulations of the microwave impedance of
short and long Josephson junctions under a variety of excitations or drive
conditions. In the case of a short junction, the only excitation possible is
an $ac$ current drive. For a long junction it is possible to study symmetric
(the excitation field direction is the same on both sides of the junction),
and antisymmetric (the excitation field direction is opposite on both sides
of the junction) field excitations, and we consider both. We also examine
the response of short and long junctions in the presence of an applied $dc$
field and, in the case of a long junction, the dependence of the impedance $%
Z_{\omega }$ on junction width $d$. These conditions are experimentally
relevant and have not been analyzed in detail before. Some of the calculated
results are consistent with experiments on crystals and films of HTS, while
other new interesting effects are described which may be observable in
future experiments.

\section{\protect\smallskip NONLINEAR IMPEDANCE OF SHORT JUNCTION}

The usual resistivitely-shunted JJ (RSJ) model typically applies to the {\it %
short} junction, in which the junction width is smaller than the Josephson
penetration depth $\lambda _{J}$. In this limit the flux in the junction is
uniformly distributed.

\subsection{\protect\smallskip ac current driven RSJ}

Recently several authors have calculated the non-linear microwave impedance
of a {\it short} RSJ \cite{sridhar,Mcdonald97,Hein95,Xie96}, subjected to a
pure $ac$ drive current represented as $I_{o}\sin (\omega t)$. In this model
the Josephson junction is described by an ideal junction J shunted by a
resistance R and capacitance C to form a parallel circuit. The total current 
$I$, flowing through the circuit has contributions from supercurrent $I_{S}$%
, normal current $I_{N}$ and displacement current $I_{D}$. While the normal
current is caused by the flow of normal electrons in the barrier while the
displacement current is due to the time varying electric field across the
junction. Therefore, for the total current $I$ one can write

\begin{equation}
I=I_{S}+I_{N}+I_{D}=I_{c}\sin (\phi )+\frac{\Phi _{0}}{2\pi R}\frac{d}{dt}%
\phi +\frac{\Phi _{0}C}{2\pi }\frac{d^{2}}{dt^{2}}\phi \text{.}
\end{equation}

\smallskip \noindent where, $\phi $ is the junction phase and $I_{c}$ is the
critical current of the junction. $\Phi _{0}$ is the unit of a flux quantum,
and the junction plasma frequency, $\omega _{p}$ is given by $\sqrt{2\pi
I_{c}/\Phi _{0}C}$. For low frequency drive currents ($\omega <<\omega
_{p}\sim 100GHz)$ the above equation can be written as,

\begin{equation}
\beta ^{-1}d\phi /dt+\sin (\phi )=i_{\omega }\sin (\omega t).
\end{equation}

\smallskip \noindent Here, $\beta =2eI_{c}R/\hbar $, $R$ is the shunt
resistance and, $i_{\omega }=I_{o}/I_{c}$. It is convenient to do the
analysis in dimensionless form

\begin{center}
\begin{equation}
d\phi /d\tau +\sin (\phi )=i_{\omega }\sin (\Omega \tau )\text{.}
\end{equation}
\end{center}

\noindent where $\tau =\beta t$, and $\Omega =\omega /\beta $. The
nonlinearity arises from the $\sin (\phi )$ term in Eq. 1. Eq. 1 is solved
numerically for $\phi $. The {\em ac} resistance, $R_{\omega }$ and
reactance $X_{\omega }$ are calculated from the relation\cite{Zhai97}

\begin{center}
\begin{equation}
\ Z_{\omega }=R_{\omega }+jX_{\omega }=\left( \frac{\omega }{2\pi i_{\omega }%
}\right) \int_{0}^{2\pi /\omega }\stackrel{.}{\phi }e^{j\omega t}dt\text{.}
\end{equation}
\end{center}

Fig. 1, shows the normalized {\em ac} resistance and reactance versus
microwave drive current $i_{\omega }$ for different $\Omega =0.01,0.1,1$. As
can be seen from the Fig. 1(a), when $\Omega $ is small, the {\em ac}
resistance shows several characteristic features: a threshold field and a
subsequent parabolic rise, above the threshold, with step-like structures.
The threshold current where a sudden increase in $R_{\omega }$ starts is
defined as critical current $i_{\omega c}$. The step height decreases with
increasing $i_{\omega }$. The reactance directly shows dynamic phase slips
occurring as the drive current is increased(Fig. 1(b)). The sharp steps
occur at values of $i_{\omega }$ for which there are bifurcations in the
solution Eq. 3, {\em i.e.} $2\pi $ phase slips in the gauge-invariant phase
difference $\phi $ across the junction. These bifurcations can cause sudden
changes in V(t) during one cycle of the alternating current and hence result
in steps\cite{cmt1}. When $\Omega $ becomes large, both the threshold and
steps in R$_{\omega }$ disappear, and the amplitude of the phase slip in $%
X_{\omega }$ becomes larger. Also, the phase slip moves to larger values of $%
i_{\omega }$ and the spacing between bifurcations becomes larger which
results in $R_{\omega }$ and $X_{\omega }$ approaching $R$ (Fig. 1).

\section{\protect\smallskip NONLINEAR IMPEDANCE OF LONG JUNCTION}

A Josephson junction is considered long when the junction width is larger
than the Josephson penetration depth, $\lambda _{J}$. In this case, for
applied magnetic field more than the junction critical field, $%
H_{c1J}=\left( \Phi _{0}/4\pi \lambda _{J}^{2}\right) \ln \left( \lambda
_{J}/t_{i}\right) $ \cite{tinkham,clem88}, the field enters the junction in
the form of Josephson fluxons. $\lambda _{J}=\sqrt{\Phi _{0}/[2\pi \mu
_{o}J_{o}(2\lambda _{L}+t_{i})]}$ is the Josephson penetration depth, and $%
\lambda _{L}$ is the London penetration depth. The RSJ model, therefore,
cannot give an adequate description of the ac or dc field responses. Fig. 2
gives a schematic representation of a long junction for both symmetric
excitation, when the microwave field $b_{\omega }$ is applied symmetrically
on either side of the junction, in which case the induced current is
anti-symmetric, and antisymmetric excitation due to the ac field $b_{\omega
} $, whence the microwave drive current flows symmetrically through the
junction.

The relation between the phase difference across the junction and the
magnetic field along the junction is given by $B_{z}\left( x,t\right) =-%
\frac{\Phi _{0}}{2\pi \left( 2\lambda _{L}+t_{i}\right) }\frac{\partial \phi
\left( x,t\right) }{\partial x}$, the relation between $\phi $ and electric
field across the junction is $E_{y}\left( x,t\right) =\frac{\Phi _{0}}{2\pi
t_{i}}\frac{\partial }{\partial t}\phi \left( x,t\right) $ and the current
density is $J_{y}\left( x,t\right) =J_{c}\sin \phi +E_{y}/\rho _{i}$. Here,
the first term represents supercurrent density and the second term normal
leakage current density, $J_{y}\left( x,t\right) $ is related to $B_{z}$ and 
$E_{y}$ by Ampere's law with a displacement current. By combining the above
relations, for frequencies smaller than the plasma frequency, the model that
represents the junction can be written as\cite{Mcdonald97}.

\begin{center}
\begin{equation}
\lambda _{J}^{2}\frac{\partial ^{2}\phi }{\partial x^{2}}-\beta ^{-1}\frac{%
\partial \phi }{\partial t}=\sin \phi \text{.}
\end{equation}
\end{center}

\noindent $\beta =2\pi \rho _{i}t_{i}J_{c}/\phi _{o}$ is a damping
parameter. Here, $t_{i}$ is the junction thickness, $\rho _{i}$ and $%
\epsilon $ are resistivity and permittivity of the junction material, $J_{c}$
is the critical current density of the junction, and $\phi _{o}$ is the flux
quantum. In dimensionless form the equation is written as

\begin{center}
\begin{equation}
\frac{\partial ^{2}\phi }{\partial X^{2}}-\frac{\partial \phi }{\partial
\tau }=\sin \phi \text{.}
\end{equation}
\end{center}

\noindent where $X=x/\lambda _{J}$, and $\tau =\beta t$.

The detailed study of the long junction depends on the parameter values and
on the boundary conditions, which will be introduced in the following for
the particuler problem considered. When the the applied microwave field $%
B_{\omega }=B_{o}\sin \omega t$ is along $\widehat{z}$-axis and junction
width infinite, the boundary conditions at $x=0$ and $x=d$ are 
\begin{equation}
\left[ \frac{\partial \phi }{\partial x}\right] _{x=0}=-b_{\omega }\sin
(\Omega \tau )\text{, }\left[ \frac{\partial }{\partial x}\phi \right]
_{x=d\rightarrow \infty }=0\text{.}
\end{equation}

Where, $b_{\omega }=2\pi B_{o}(2\lambda _{L}+t_{i})/\phi _{o}$, $\Omega
=\omega /\beta $. Eq. 7 is solved numerically for $\phi $ and, as in the
case of small junction. Using Eq. 4 and boundary conditions the expressions
for $R_{\omega }$ and $X_{\omega }$ are obtained.

Fig. 3 shows $R_{\omega }/R_{n}$ and $X_{\omega }/R_{n}$ versus $b_{\omega }$
for different values of $\Omega =0.01$, $0.1$ and $1$, for an infinite long
junction with $R_{n}=\sqrt{\mu _{o}\rho _{i}\omega (2\lambda
_{L}+t_{i})/(2t_{i})}$. When $\Omega $ is small, the variation in $R_{\omega
}$ is similar to the RSJ model calculation, in that it has a threshold and a
parabolic rise with step-like structures. However, no clear phase slips are
observed in $X_{\omega }$ versus $b_{\omega }$. When $\Omega $ is large, at
the threshold the jump becomes higher and for further increase in $b_{\omega
}$ the steps become smooth. When $\Omega $ is very large the steps are
washed out. The {\em ac} reactance $X_{\omega }$ also shows a similar
transitions from a step-like response to a smooth curve.Indeed the overall
behavior of $R_{\omega }$ and $X_{\omega }$ are similar, this is essentially
due to the ``bulk'' nature of the configuration. In the long junction the ac
impedance is associated with the vortex - antivortex propagations. For short
junctions the notion of a vortex - antivortex pair does not appear because
the Josephson vortex has a finite size of the order $2J_{c}\lambda _{J}$.
Therefore, the physical mechanism leading to step structures in short and
long junctions are different. For long junctions the steps are due to
synchronous coupling of the oscillating vortex or antivortex with the ac
current \cite{chang86}. Essentially, the steps in Fig. 3 occur due to the
changes in the number of vortices in the junction and the step height is
dependent on the microwave power. For higher values of $\Omega $ vortex
entry takes place at higher field and hence steps become broader and are
moved to larger values of $b_{\omega }$.

In the limit the junction width $d\thicksim \lambda _{J}$, the boundary
condition $\left[ \frac{\partial \phi }{\partial x}\right] _{x=d}\neq 0$.
This problem can be studied in two ways, the symmetric and antisymmetric
cases (see Fig. 2).

\subsection{\protect\smallskip Symmetric field excitation (antisymmetric
current drive)}

When a microwave field is applied on either side of the junction ( Fig. 2a),
then this corresponds to a symmetric microwave excitation. The boundary
condition in this case is 
\begin{equation}
\left[ \frac{\partial }{\partial x}\phi \right] _{x=0,d}=-b_{\omega }\sin
(\Omega \tau )\text{.}
\end{equation}
In Fig. 4 $R_{\omega }/R_{n}$ versus $b_{\omega }$ responses for three
different values of $\Omega $ and four different values of the width, $%
d/\lambda _{J}=5$, $10$, $50$ and $\infty $, are shown. When $\Omega $ is
small ( Fig.4a), $R_{\omega }$ is small for low values of $d/\lambda _{J}$
compared to the high values of the latter. $R_{\omega }$ approaches $R$ at
high values of $b_{\omega }$ for infinite junction width. The threshold and
the steps can be seen clearly for all values of $d/\lambda _{J}$. On the
contrary, when $\Omega $ is large (Fig. 4c), it can be seen that there is
almost no effect on $R_{\omega }$ with the change in the junction width,
except when junction width is too small.

Both at high field ($b_{\omega }\rightarrow \infty $) and large damping ($%
\Omega \sim 1$)the junction behaves like a normal metal with $R_{\omega }$
approaching its asymptotic value $R_{n}$. The result can be understood by
considering the fact that at these two conditions, the $R_{\omega }$
response is dominated by loss arising from normal current than the
supercurrent. The manifestation of dominance of normal contribution can be
seen in Fig. 4(c) in two ways. Firstly, the $R_{\omega }$ response at larger
damping is smooth sans any threshold and step like features. Secondly, since
the normal current contribution to $R_{\omega }$ depends on the Josephson
skin depth, which is invariant for any fixed $\Omega ,$ variation in the
junction width does not have much affect on $R_{\omega }$ , and therefore,
results in the same change in the magnitude of $R_{\omega }$ at various $%
d/\lambda _{J}$ values.

However, for all the values of $\Omega $ when $d/\lambda _{J}$ is too small $%
(d/\lambda _{J}=2)$, it can be seen that $R_{\omega }$ is small. In fact, as 
$d/\lambda _{J}$ approaches zero, the junction effects should not be
observed as the total current gets canceled out due to asymmetry on either
sides of the junction.

\subsection{{\bf \protect\smallskip }Antisymmetric field excitation
(symmetric current drive)}

If an $ac$ current is passed through the junction ( Fig. 2b), then the
direction of the induced magnetic field is opposite on both sides of the
junction making this configuration an antisymmetric field-driven case. The
corresponding boundary conditions are

\begin{equation}
\left[ \frac{\partial }{\partial x}\phi \right] _{x=0}=-b_{\omega }\sin
(\Omega \tau )\text{, }\left[ \frac{\partial }{\partial x}\phi \right]
_{x=d}=b_{\omega }\sin (\Omega \tau )\text{.}
\end{equation}
In Fig. 5 plots of $R_{\omega }/R_{n}$ versus $b_{\omega }$ for $d/\lambda
_{J}=5$, $10$, $20$ and $\infty $, at three different values of $\Omega $,
are shown. When $\Omega $ is small, compared to the infinite width case
(Fig. 3), $R_{\omega }$ is larger for small widths. It is worth noting here
that this response is quite opposite to the symmetric field excitation,
discussed in the earlier section, where $R_{\omega }$ is larger when the
width is large. When $\Omega $ is large enough ($\Omega =1$), as in the
symmetric case, the change in width does not show much affect on $R_{\omega
} $, unless the width is too small. The interesting case, here, is when $%
\Omega $ is in the middle region ($\Omega =0.1$, Fig. 5b). In this case for
junction width several times the Josephson penetration depth ($d/\lambda
_{J}=5$), the $R_{\omega }$ response shows a threshold and rises
parabolically with the step-like structures for increasing $b_{\omega }$. It
is similar to the result of short junction RSJ calculation and long junction
small $\Omega $ case. As the junction width increases a few times more ($%
d/\lambda _{J}=10$), $R_{\omega }$ shows sudden jumps as $b_{\omega }$
increases. This jumpiness disappears for higher values of junction width,
and eventually, when $d/\lambda _{J}=\infty $ the steps become broader and
smooth.

These results can be explained by invoking the fact that in the
antisymmetric field excitation since the currents on both sides of the
junction flow in the same direction, as $d/\lambda _{J}$ decreases current
density increases, for any particular field. Such a current crowding causes $%
R_{\omega }$ to increase for smaller values of $d/\lambda _{J}.$ Also, the
difference between the step patterns in the symmetric and antisymmetric
field excitations is due to different vortex or antivortex interactions with
ac current, as they move across the junction width.

\section{NONLINEAR IMPEDANCE OF $dc$ BIASED (CURRENT OR FIELD) JUNCTION}

An important class of experiments is where the microwave absorption is
measured in the presence of a dc magnetic field which is varied. For this
reason, we consider below the case of a JJ which is biased by a dc current
or magnetic field. Naturally, the short junction is biased by a dc current,
and the infinitely long junction by a dc magnetic field ( we consider the
case of symmetric excitation only), and we calculate the ac impedance as a
function of the dc bias.

\subsection{Short junction :\ dc current biased, ac current driven}

If a dc current, $I_{dc}$ is superposed on $i_{\omega }$, the driving
current becomes $I_{o}\sin (\omega t)+I_{dc}$. The corresponding junction
equation can then be written as

\begin{equation}
d\phi /d\tau +\sin (\phi )=i_{\omega }\sin (\Omega \tau )+i_{dc}\text{.}
\end{equation}

\noindent where $i_{dc}=I_{dc}/I_{c}$. In Fig. 6, we plot $R_{\omega }$ and $%
X_{\omega }$ versus $i_{\omega }$ for various values of $i_{dc}=0$, $0.5$, $%
2 $, and $10$ at fixed $\Omega =0.1$. There are two regimes of behavior
noticeable, corresponding to $i_{dc}>i_{\omega c}$ or $i_{dc}<i_{\omega c}$.

When $i_{dc}=0$, the critical current $i_{\omega c}(i_{dc}=0)$ is found to
be $1.1$(Fig. 6). When the $dc$ drive current $i_{dc}$ is smaller than the $%
zero$ $dc$ critical current (i.e. $0<$ $i_{dc}<1.1$), compared to the
response of $R_{\omega }$ versus $i_{\omega }$ for $i_{dc}=0$, it can be
seen that the critical current $i_{\omega c}$ decreases and that both the
step size and step rising pattern are changed. In Fig. 7, we plot the
critical current $i_{\omega c}$ as a function of $i_{dc}$ for $\Omega =0.1$.
A linear fit $i_{\omega c}=ai_{dc}+b$ yields $a\thickapprox -1$ and $%
b\thickapprox 1.1$. Note however that this critical current value is weakly
frequency-dependent.

When $i_{dc}$ is larger than the $zero$ $dc$ critical current (i.e. $%
i_{dc}>1.1$), initially apparent jumps appear and with further increase of
the $ac$ drive current $i_{\omega }$, $R_{\omega }$ begins to decrease with
sudden intermittent steps and eventually approaches the $R_{\omega
}(i_{\omega }\rightarrow \infty ,$ $i_{dc}=0)$. This regime $%
i_{dc},i_{\omega }>i_{\omega c}$ is very interesting as it appears to show
strong phase-locking effects in the ac impedance, similar to the Shapiro
steps seen in dc $I-V$ characteristics.

For very large values of $i_{dc}$ ($i_{dc}=10$), as expected, the effect of
the microwave current is negligible and the {\em ac} resistance approaches
the shunt resistance $R$.

In the case of {\em ac} reactance $X_{\omega }$, which is plotted in Fig. 6
bottom panel, the phase slip is not affected by the increase in the $dc$
current $i_{dc}$, however, the amplitude of phase slip reduces dramatically
for larger values of $i_{dc}$.

In Fig. 8, we plot $R_{\omega }$ and $X_{\omega }$ versus $i_{dc}$ for
different microwave drive currents $i_{\omega }=0.5$, $1$, $2$, $10$. It can
be observed from the figure that for all values of $i_{\omega }$, as $i_{dc}$
increases, $R_{\omega }$ initially increases, reaches a maximum and
approaches $R$ eventually. On the other hand, X$_{\omega }$ decreases
initially with the increase in $i_{dc}$. The occurrence of phase slip can be
seen clearly when $i_{dc}$ is small. Large values of either $i_{\omega }$ or 
$i_{dc}$ diminishes the effect of phase slip.

\subsection{Long Junction : dc magnetic field biased, ac field driven}

The presence of dc magnetic field, $B_{dc}$ is studied for infinite long
junction because in this case the $R_{\omega }$ response is same in both
symmetric and antisymmetric excitations, as can be seen in Figs. 4 and 5.
The calculations are, however, done for symmetric excitaion. When $B_{dc}$
is applied in addition to the microwave field, the boundary condition
becomes 
\begin{equation}
\left[ \frac{\partial }{\partial x}\phi \right] _{x=0}=-b_{\omega }\sin
(\omega t)-b_{dc}\text{.}
\end{equation}
where, $b_{dc}=2\pi B_{dc}(2\lambda _{L}+t_{i})/\phi _{o}$. Considering $%
\Omega =0.1$ and $d/\lambda _{J}=\infty $ case, the $R_{\omega }/R_{n}$
versus $b_{\omega }$ response at various $b_{dc}=0$, $0.1$, $1$ are shown in
Fig. 9. In Fig. 10 the plots of $R_{\omega }/R_{n}$ versus $b_{dc}$ for two
different $b_{\omega }$ values, $b_{\omega }=0.1$ and $1$, are shown. It can
be seen in Fig. 9 that when the dc field is high, $R_{\omega }$ at low ac
field is high and as $b_{\omega }$ increses it decreases and reaches the
asymptotic value $R_{\omega }$ when $b_{\omega }\rightarrow b_{dc}$. This
high ac resistance is due to the loss arising from vortex flow whose
response is dependent on the intrinsic pinning in the junction. The same
mechanism results in $R_{\omega }$ increasing beyond $R_{n}$ as the dc field
is increased.

A very intriguing result is the case for intermediate $b_{\omega }=0.1$,
shown in Fig. 10, for the parameter $\Omega =0.1$. The field dependence is
strikingly regular, almost periodic. As noted below, this is similar to
experimental data seen in microwave absorption experiments on YBCO single
crystals.

\section{REMARKS}

The recent observation that bulk single crystals of high temperature
superconductors behave like Josephson junctions in their microwave response,
and are even of higher quality than fabricated junction \cite{Zhai97}, was
the original motivation for this work. In Fig. 11 the experimental microwave
surface resistance variation as a function of microwave field $\left(
H_{rf}\right) $ recorded on a high quality single crystal of $%
YBa_{2}Cu_{3}O_{7-\delta },$ at temperature of $3K$, is shown. It is
important to note that the $R_{s}$ variation mimics a single long junction
response for symmetric excitation (see Fig.3), with $\Omega =0.03$. Since
the measuring frequency is $10GHz$, this gives a damping frequency of $%
330GHz $ equivalent to an $I_{c}R=0.8\,meV$. A detailed comparison of
experimental data with simulations will be presented elsewhere.

Very interesting results which can be attributed to JJ's have been observed
in BSCCO \cite{Jacobs96,Jacobs96b}. There the microwave surface resistance
shows sudden jumps or transitions between apparently quantized values when
the microwave field $H_{\omega }||\,\hat{c}$-axis. However, it is probably
necessary here to also include the inertial term even in the RSJ model,
which leads to a very rich dynamics and apparently chaotic behavior.

Another class of experiments which have been analyzed using an RSJ model is
the Modulated Microwave Absorption experiments \cite{MAMMA}, in which the
microwave absorption is measured in the presence of a dc bias field, along
with a very low frequency ($<100Hz$) modulation field. Here again the
properties calculated were in terms of dc quantities, and not the true
microwave impedance as is done here, and hence the analysis should be redone
along the lines of the present paper, as we have in the previous section.

Indeed, as Fig. 10 shows, a striking nearly periodic behavior on the
microwave bias field can be observed in the caculations. Similar periodic
behavior has often been observed in modulation experiments on single
crystals \cite{periodic}. Our calculations shows that this type of behavior
can occur in long junctions for a judicious choice of the parameters $\Omega 
$ and $b_{\omega }$.

A surprising theme that emerges from the experiments mentioned above, when
compared with calculations as presented in this paper, is that the
experimental data can often be well described by the response of a single
junction, rather than the multitude of ``weak links'' one would expect due
to inhomogeneities. Thus the experiments and analysis taken together
strongly suggest the presence of a single periodic ($\sin (\phi )$) response
of the superconductor to applied magnetic fields. One can imagine that this
arises from a single dominant weak link in the system, or of many weak links
acting coherently. As pointed out in ref. \cite{Zhai97}, the degree of
coherence is striking and strongly suggests a fundamental mechanism, such as
the possibility of bulk coupled superconductors.

Previously, several authors have modeled Josephson junction response in
granular superconductors in terms of strongly coupled and weakly coupled
Josephson junctions. Experimentally, it is found that HTS of different
families often respond to microwave excitations in a similar manner,
including ceramics and films and different materials like YBCO and BSCCO.
Hylton et al.\cite{hylton} have used a resistively shunted junction (RSJ)
model to describe the residual microwave losses in thin films. In this model
the grains are considered to be purely inductive and the grain boundary weak
links the RSJ's. This model has later been expanded and used by various
authors\cite{oates93,mahel,atta91}. However it is now clear that this early
analysis needs to be replaced by the more exact analysis such as presented
in this paper and by McDonald et. al. \cite{Mcdonald97}.

While the discussion in this paper has been entirely in the context of
superconducting Josephson junctions, we are aware that the fundamental
equations are applicable to a variety of nonlinear systems. These results
are generic to all systems with a $\cos (\phi )$ potential in the overdamped
limit subjected to an ac drive. Hence the results presented here have a
general validity beyond superconductivity, wherever the ac impedance is a
significant quantity which is experimentally accessible. A particularly
interesting system is that of striped metals, where the dynamical response
is again of the Josephson type \cite{Hasselman98}, and where all the
analysis obtained here should apply .

In summary, we have presented the nonlinear {\em ac} resistance and
reactance analysis for both short and long junctions under various
conditions. In the case of a short junction, the higher damping leads to
less pronounced nonlinearities. The steps observed in the {\em ac}
resistance are due to 2$\pi $ phase slips in the gauge invariant phase
difference across the junction. The application of dc field reduces the
critical current and when the dc current is more than the ac current, phase
locking effects are seen in the ac impedance, similar to Shapiro steps in
the dc $I-V$ characteristics.

In the case of a long junction, the step-like features in {\em ac} impedance
appear due to changes in the number of fluxons. When the damping is large,
the steps are pushed to higher values of current as the vortex entry takes
place at high currents. In the symmetric and antisymmetric filed excitations
it found that, at large damping or high fields, the {\em ac} resistance
response is dominated by the resistive (normal ) contribution and hence
leads to same magnitude and variation of $R_{\omega }$, for various values
of the junction width. Overall, for both short and long junctions, in high
current and large damping limit, the {\em ac} resistance approaches the
limiting dc resistance value and the nonlinearities in its response are less
pronounced. These interesting features may be observable in experiments on
crystals and films of HTS and other superconductors.

\section{ACKNOWLEDGMENTS}

Work at Northeastern University was supported by NSF-ECS-9711910.

\section{Figure Captions}

%TCIMACRO{
%\TeXButton{figure1}{\begin{figure}
%\caption{ RSJ model calculation of  ac $\left( a\right) $ resistance and 
% $\left( b\right) $ reactance versus $i_{\omega}$ of a short junction for three 
%different values of $\Omega ,$ 
%the damping parameter. The step like features seen in $\ R_{\omega}$ are due to 2$\pi $ 
%phase slips in the phase difference across the junction.  }
%\label{fig1}
%\end{figure}}}
%BeginExpansion
\begin{figure}
\caption{ RSJ model calculation of  ac $\left( a\right) $ resistance and 
 $\left( b\right) $ reactance versus $i_{\omega}$ of a short junction for three 
different values of $\Omega ,$ 
the damping parameter. The step like features seen in $\ R_{\omega}$ are due to 2$\pi $ 
phase slips in the phase difference across the junction.  }
\label{fig1}
\end{figure}%
%EndExpansion
\label{fig1}\label{figure1}

%TCIMACRO{
%\TeXButton{figure2}{\begin{figure}
%\caption{ Cross sectional view of long Josephson junction for $\left( a\right) $  symmetric 
%field excitation - antisymmetric current drive,  and $\left( b\right) $ antisymmetric field
% excitation - symmetric curret drive. }
%\label{fig2 }
%\end{figure}}}
%BeginExpansion
\begin{figure}
\caption{ Cross sectional view of long Josephson junction for $\left( a\right) $  symmetric 
field excitation - antisymmetric current drive,  and $\left( b\right) $ antisymmetric field
 excitation - symmetric curret drive. }
\label{fig2 }
\end{figure}%
%EndExpansion

%TCIMACRO{
%\TeXButton{figure3}{\begin{figure}
%\caption{ $\left( a\right) $ $R_{\omega}$ and $\left( b\right) $ $X_{\omega}$ versus $i_{\omega}$ for
% infinite long junction for various values of $\Omega $. The step like structures are due 
%to changes in the number of vortices in the junction. At higher values $\Omega $ vortex
% nucleation takes place at higher currents
% and hence the steps are pushed to higher current values.  }
%\label{fig3}
%\end{figure}}}
%BeginExpansion
\begin{figure}
\caption{ $\left( a\right) $ $R_{\omega}$ and $\left( b\right) $ $X_{\omega}$ versus $i_{\omega}$ for
 infinite long junction for various values of $\Omega $. The step like structures are due 
to changes in the number of vortices in the junction. At higher values $\Omega $ vortex
 nucleation takes place at higher currents
 and hence the steps are pushed to higher current values.  }
\label{fig3}
\end{figure}%
%EndExpansion

%TCIMACRO{
%\TeXButton{figure4}{\begin{figure}
%\caption{ $R_{\omega}$ versus $b_{\omega}$ for symmetric field excitation when
% $\Omega =0.01$ $\left( a\right) ,0.1$ $\left( b\right) ,$ 
%and $1.0$ $\left( c\right) $, and for various  junction widths. When the damping is  high
%and for higher currents the resistive response dominates and hence threshold 
%step like features disappear in the 
%$R_{\omega}$ response (bottom panel).  }
%\label{fig4}
%\end{figure}}}
%BeginExpansion
\begin{figure}
\caption{ $R_{\omega}$ versus $b_{\omega}$ for symmetric field excitation when
 $\Omega =0.01$ $\left( a\right) ,0.1$ $\left( b\right) ,$ 
and $1.0$ $\left( c\right) $, and for various  junction widths. When the damping is  high
and for higher currents the resistive response dominates and hence threshold 
step like features disappear in the 
$R_{\omega}$ response (bottom panel).  }
\label{fig4}
\end{figure}%
%EndExpansion

%TCIMACRO{
%\TeXButton{figure5}{\begin{figure}
%\caption{ $R_{\omega}$ versus $b_{\omega}$ for antisymmetric field excitation when
% $\left( a\right) $ $\Omega = 0.01$,  $\left( b\right) $ $0.1$ 
%and $\left( c\right) $ $1.0$, and for varying junction widths. }
%\label{fig5}
%\end{figure}}}
%BeginExpansion
\begin{figure}
\caption{ $R_{\omega}$ versus $b_{\omega}$ for antisymmetric field excitation when
 $\left( a\right) $ $\Omega = 0.01$,  $\left( b\right) $ $0.1$ 
and $\left( c\right) $ $1.0$, and for varying junction widths. }
\label{fig5}
\end{figure}%
%EndExpansion

%TCIMACRO{
%\TeXButton{figure6}{\begin{figure}
%\caption{$R_{\omega}$ $\left( a\right) $ and $X_{\omega}$ $\left( b\right) $ versus 
%$i_{\omega}$ of short junction in the presence
%a dc current $i_{dc}$. When $i_{dc}$ is smaller than the $zero$ $dc$
% critical current, $i_{\omega c}$ ($i.e$. $0<$ $i_{dc}<1.1$), it can be seen that $i_{c}$ decreases.
% When $i_{dc}$ is larger than the $zero$ $dc$ critical current ($i.e.$ $i_{dc}>1.1$),
%$R_{\omega}$ begins to decrease with sudden intermitent steps and eventually approaches the
%the limiting resistive value. 
%This regime $i_{dc},i_{\omega}>i_{\omega c}$ is very interesting 
%as it appears to show phase-locking effects in the ac impedance, similar 
%to the Shapiro  steps seen in dc $I-V$ characteristics.  }
%\label{fig2}
%\end{figure}}}
%BeginExpansion
\begin{figure}
\caption{$R_{\omega}$ $\left( a\right) $ and $X_{\omega}$ $\left( b\right) $ versus 
$i_{\omega}$ of short junction in the presence
a dc current $i_{dc}$. When $i_{dc}$ is smaller than the $zero$ $dc$
 critical current, $i_{\omega c}$ ($i.e$. $0<$ $i_{dc}<1.1$), it can be seen that $i_{c}$ decreases.
 When $i_{dc}$ is larger than the $zero$ $dc$ critical current ($i.e.$ $i_{dc}>1.1$),
$R_{\omega}$ begins to decrease with sudden intermitent steps and eventually approaches the
the limiting resistive value. 
This regime $i_{dc},i_{\omega}>i_{\omega c}$ is very interesting 
as it appears to show phase-locking effects in the ac impedance, similar 
to the Shapiro  steps seen in dc $I-V$ characteristics.  }
\label{fig2}
\end{figure}%
%EndExpansion

%TCIMACRO{
%\TeXButton{figure7}{\begin{figure}
%\caption{ Variation of $i_{\omega c}$ with $i_{dc}.$ Note 
%that $i_{\omega c}$ decreases linearly with slope $\approx -1$.  }
%\label{fig7}
%\end{figure}}}
%BeginExpansion
\begin{figure}
\caption{ Variation of $i_{\omega c}$ with $i_{dc}.$ Note 
that $i_{\omega c}$ decreases linearly with slope $\approx -1$.  }
\label{fig7}
\end{figure}%
%EndExpansion

%TCIMACRO{
%\TeXButton{figure8}{\begin{figure}
%\caption{ $\left( a\right) $ $R_{\omega}$ and $\left( b\right) $ $X_{\omega}$ versus $i_{dc}$ 
%for various values of $i_{\omega}$. With the increase in $i_{dc},$ while $R_{\omega}$ increases 
%initially, $X_{\omega}$ decreases. The occurance of phase slip is pronounced at low ac currents.  }
%\label{fig4}
%\end{figure}}}
%BeginExpansion
\begin{figure}
\caption{ $\left( a\right) $ $R_{\omega}$ and $\left( b\right) $ $X_{\omega}$ versus $i_{dc}$ 
for various values of $i_{\omega}$. With the increase in $i_{dc},$ while $R_{\omega}$ increases 
initially, $X_{\omega}$ decreases. The occurance of phase slip is pronounced at low ac currents.  }
\label{fig4}
\end{figure}%
%EndExpansion

%TCIMACRO{
%\TeXButton{figure9}{\begin{figure}
%\caption{ $R_{\omega}$ dependence on $b_{\omega}$ for an infinite 
%long junction when $\Omega =0.1$ in the presence of $dc$ field.  }
%\label{fig9}
%\end{figure}}}
%BeginExpansion
\begin{figure}
\caption{ $R_{\omega}$ dependence on $b_{\omega}$ for an infinite 
long junction when $\Omega =0.1$ in the presence of $dc$ field.  }
\label{fig9}
\end{figure}%
%EndExpansion

%TCIMACRO{
%\TeXButton{figure10}{\begin{figure}
%\caption{ $R_{\omega}$ dependence on $b_{dc}$ for an
% infinite long junction when $\Omega =0.1$ in the presence of $ac$ field.  }
%\label{fig10}
%\end{figure}}}
%BeginExpansion
\begin{figure}
\caption{ $R_{\omega}$ dependence on $b_{dc}$ for an
 infinite long junction when $\Omega =0.1$ in the presence of $ac$ field.  }
\label{fig10}
\end{figure}%
%EndExpansion

\smallskip 
%TCIMACRO{
%\TeXButton{figure11}{\begin{figure}
%\caption{ Microwave surface resistance $(R_s)$ versus microwave magnetic field $(H_{rf} \| c-axis)$
%recorded at $10 GHz$ abd $3K$, of a very high quality single of $YBa_2 Cu_3 O_{7- \delta}$. Note that the $R_s$ variation  can very well be described by 
% a single Josphson junction response.}  
%\label{fig11}
%\end{figure}}}
%BeginExpansion
\begin{figure}
\caption{ Microwave surface resistance $(R_s)$ versus microwave magnetic field $(H_{rf} \| c-axis)$
recorded at $10 GHz$ abd $3K$, of a very high quality single of $YBa_2 Cu_3 O_{7- \delta}$. Note that the $R_s$ variation  can very well be described by 
 a single Josphson junction response.}  
\label{fig11}
\end{figure}%
%EndExpansion


\begin{references}
\bibitem{cmt1}  In ref. \cite{Mcdonald97}, the authors analyse the sudden
changes in v(t) for various values of $i_{\omega }$. It is found that for
two different values of $i_{\omega }$ which are close by, for any particular 
$\tau $, the voltage responses differ from each other by a change in the
sign of the voltage pulses. Therefore, the sharp features in $R_{\omega }$
and $X_{\omega }$ correspond to the inversion of a voltage pulse at two
different times during one cycle of the alternating current.
\end{references}
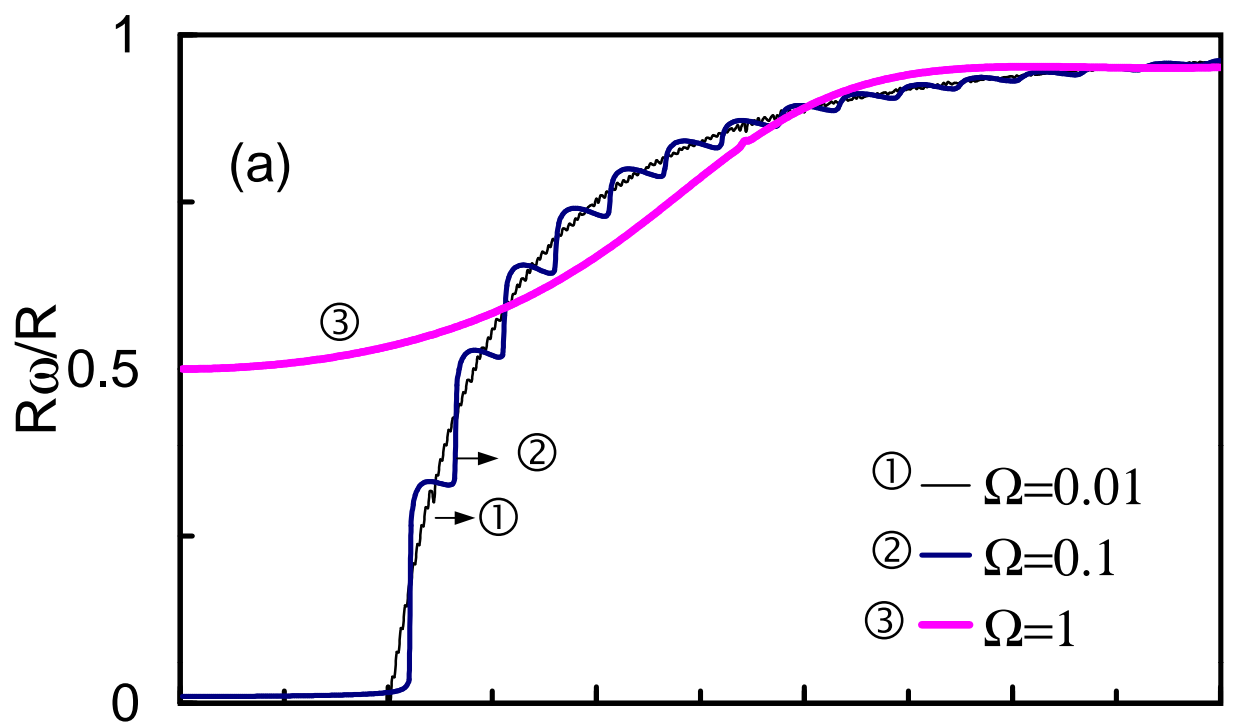
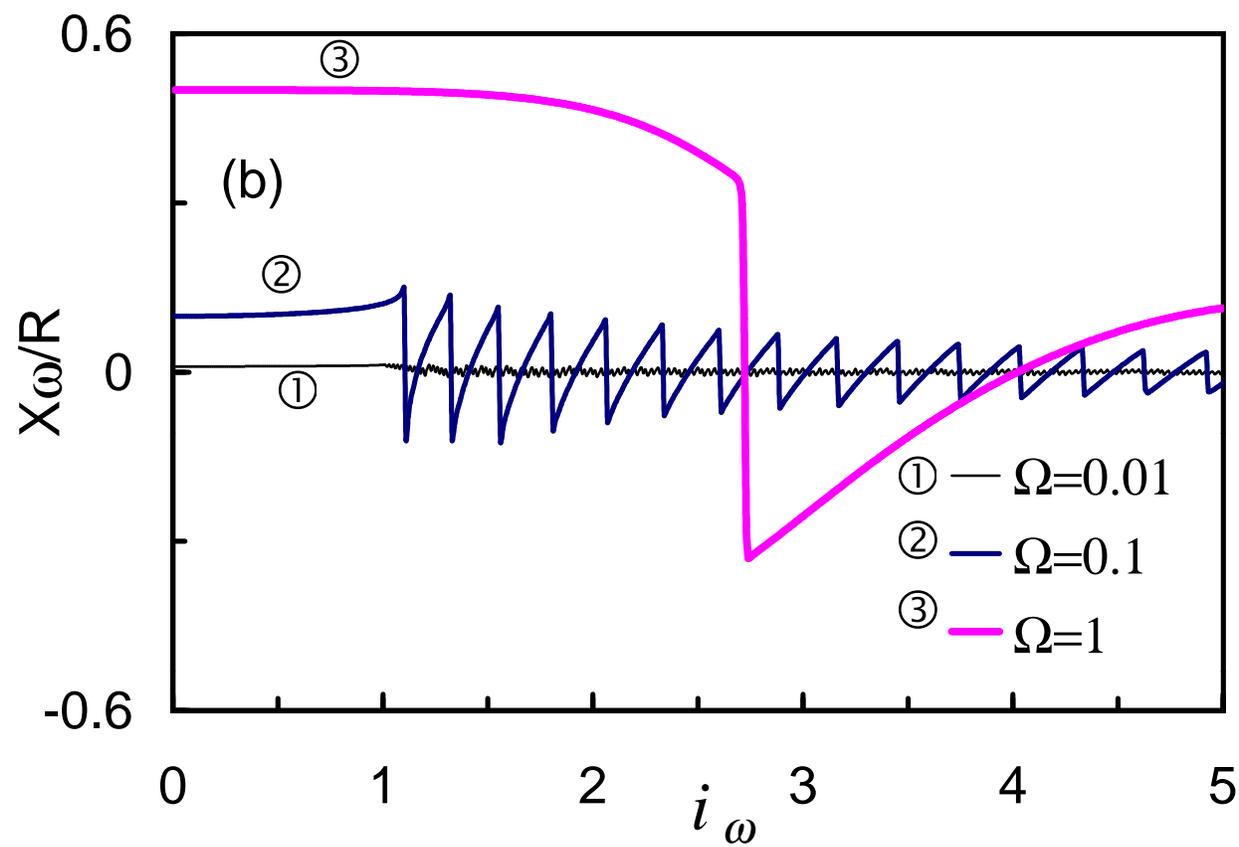
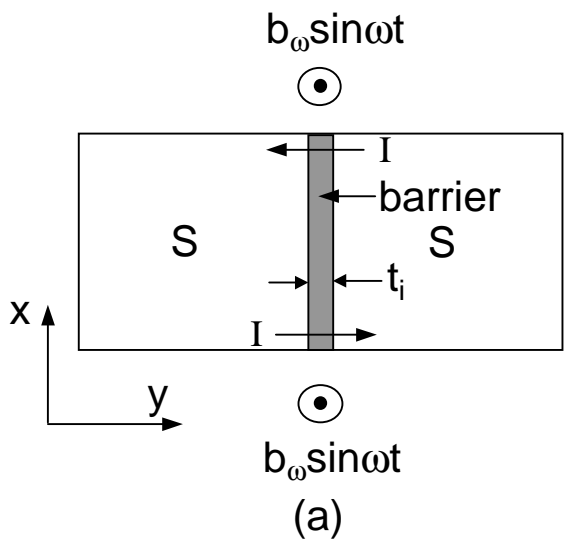
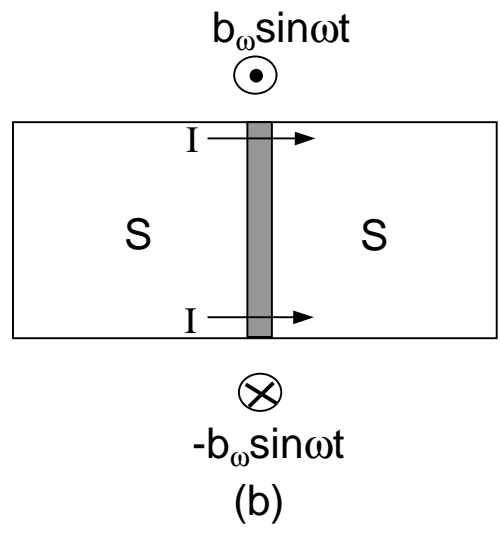
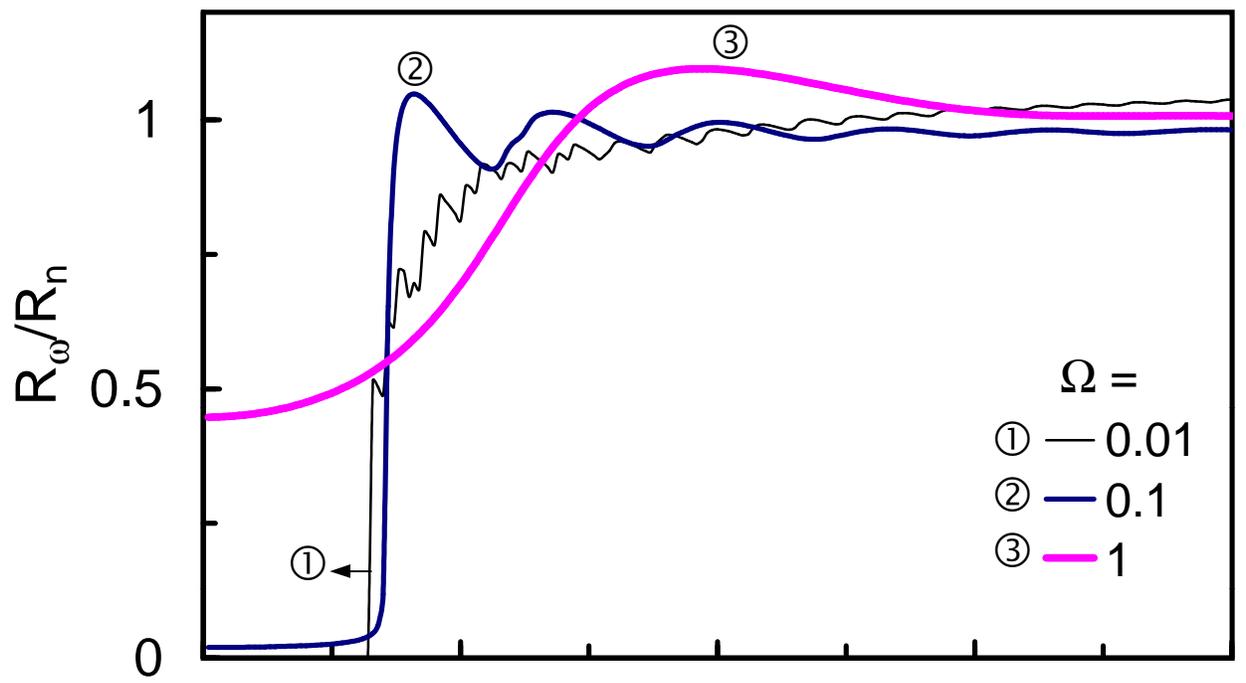
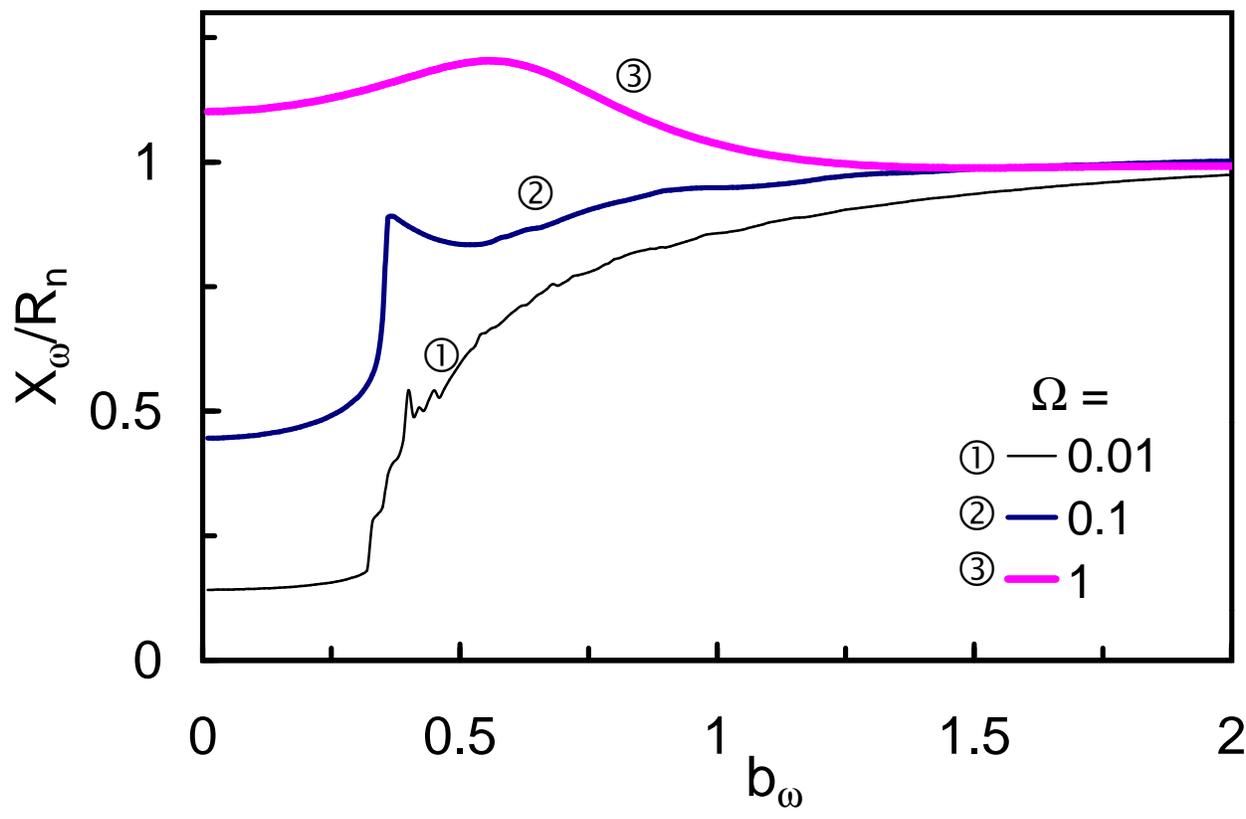
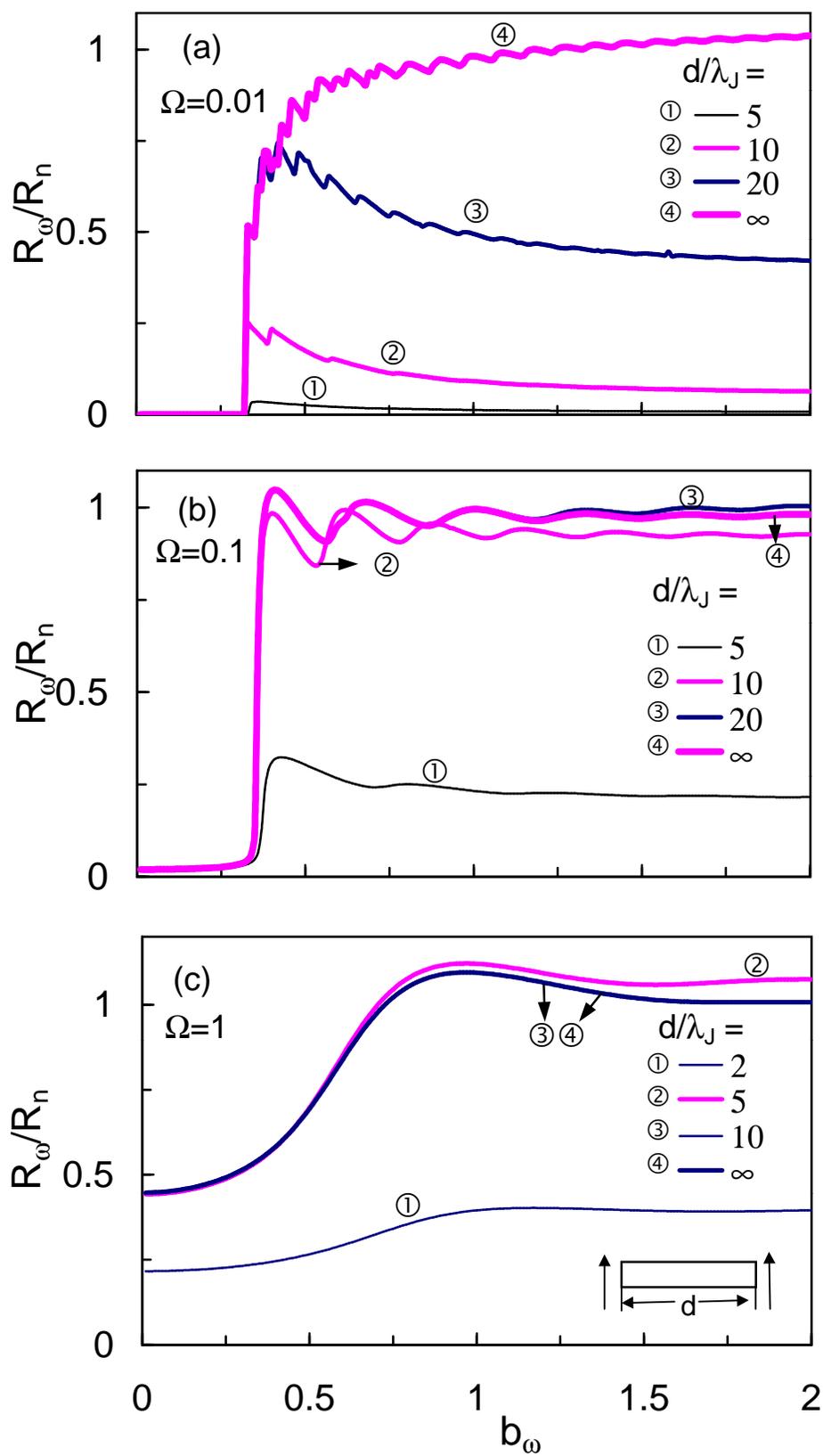
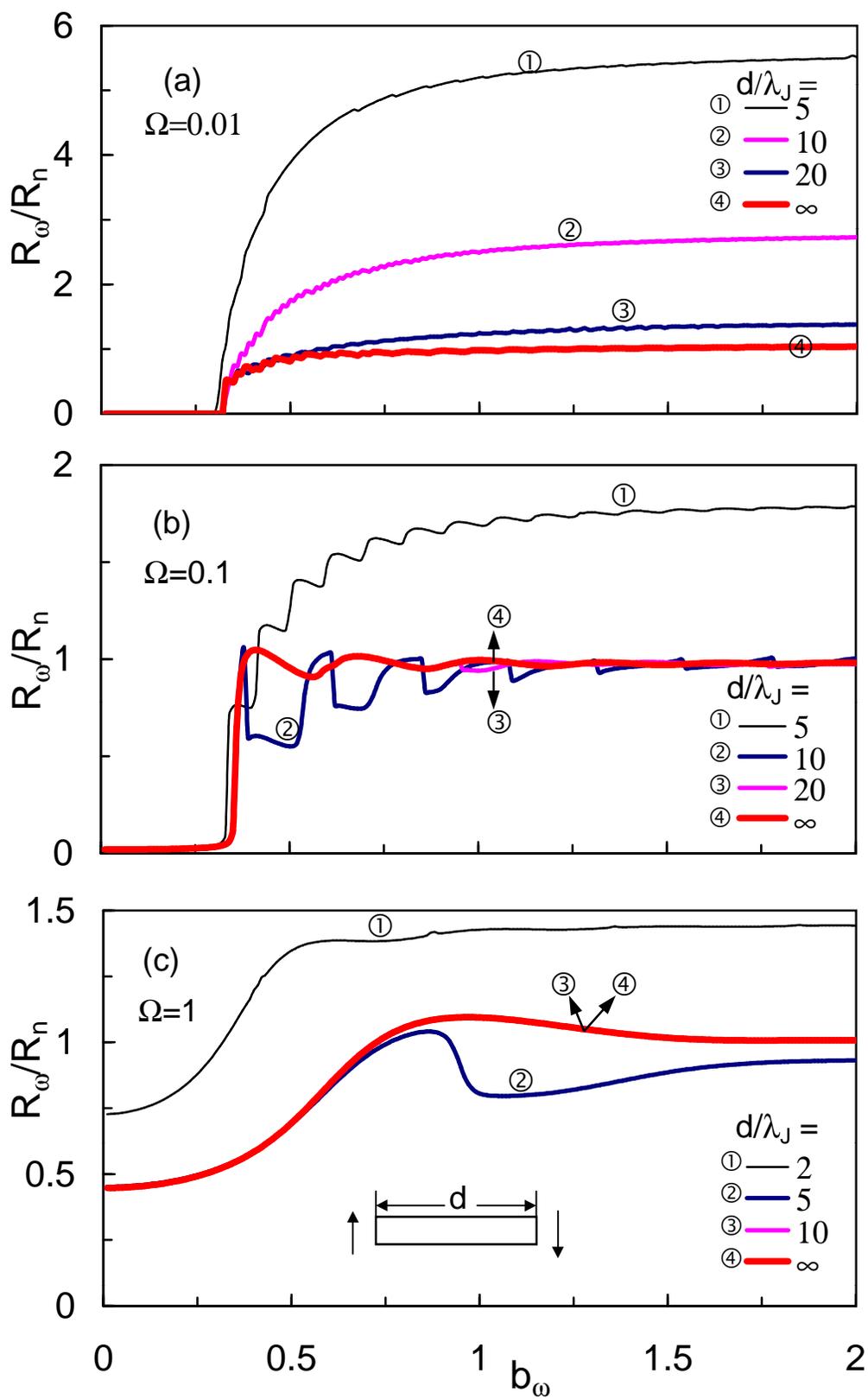
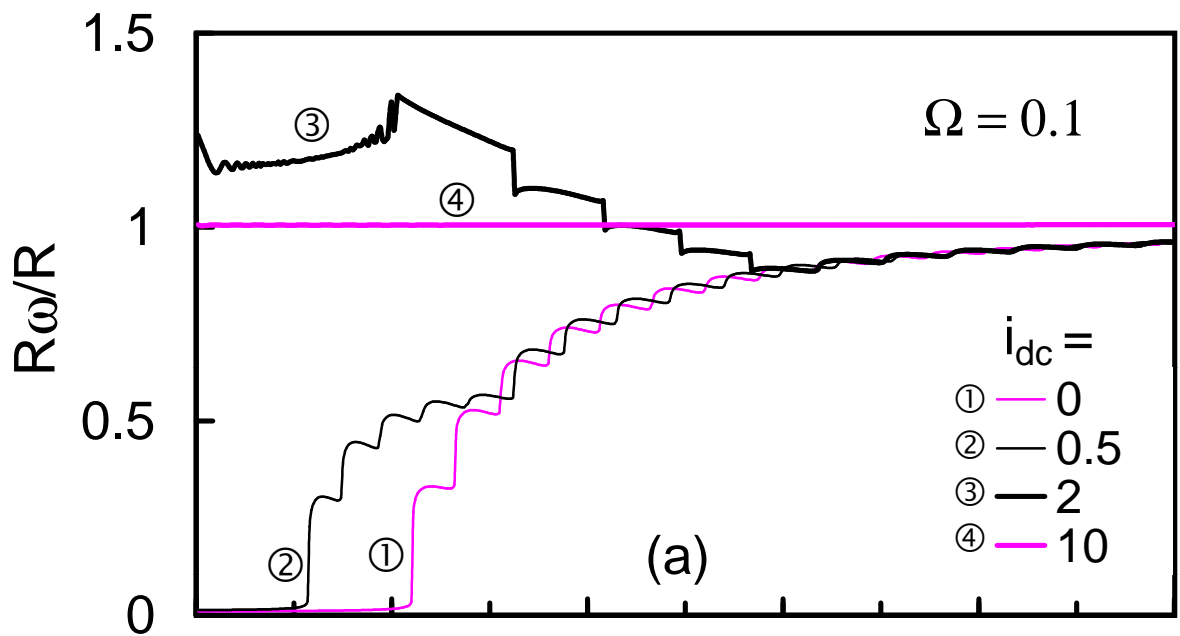
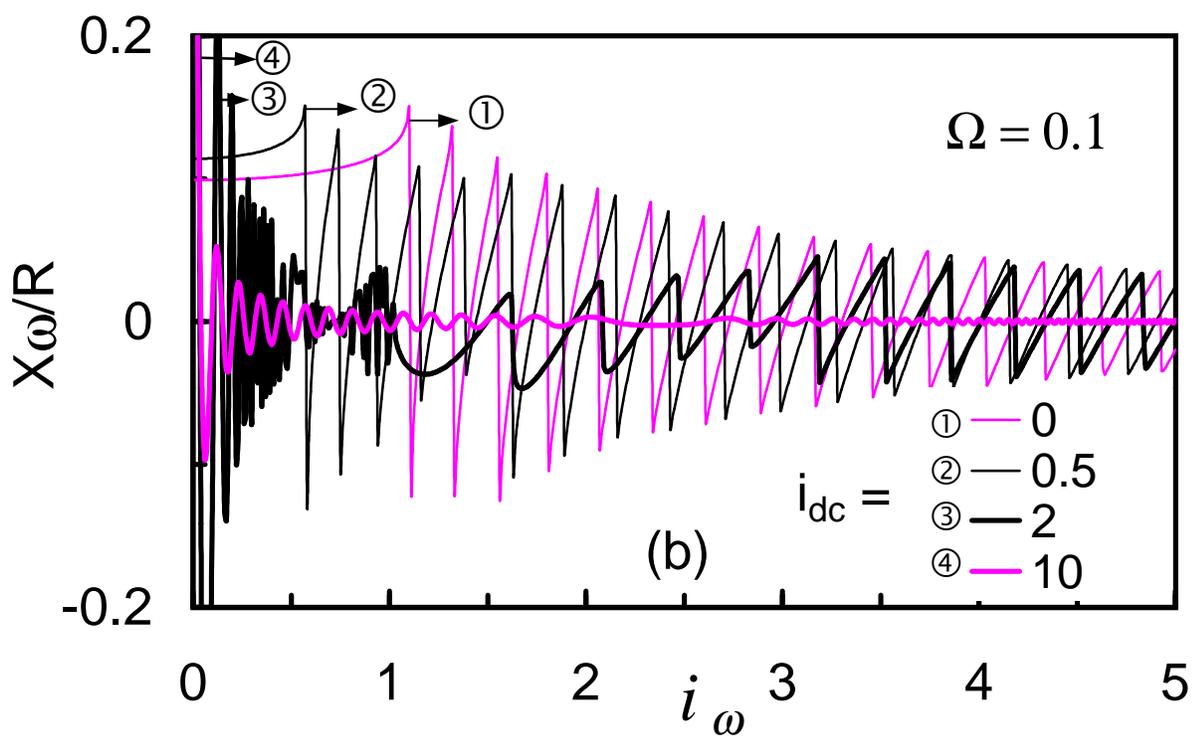
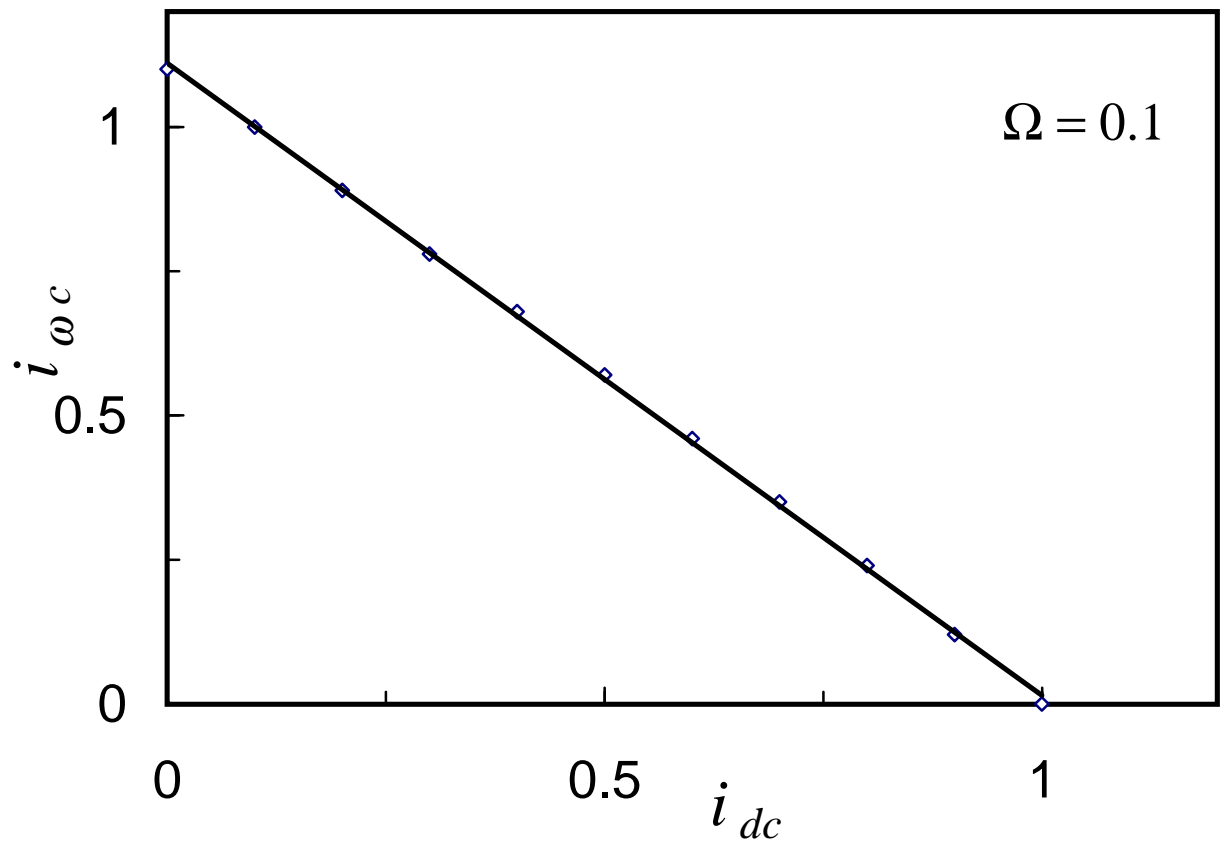
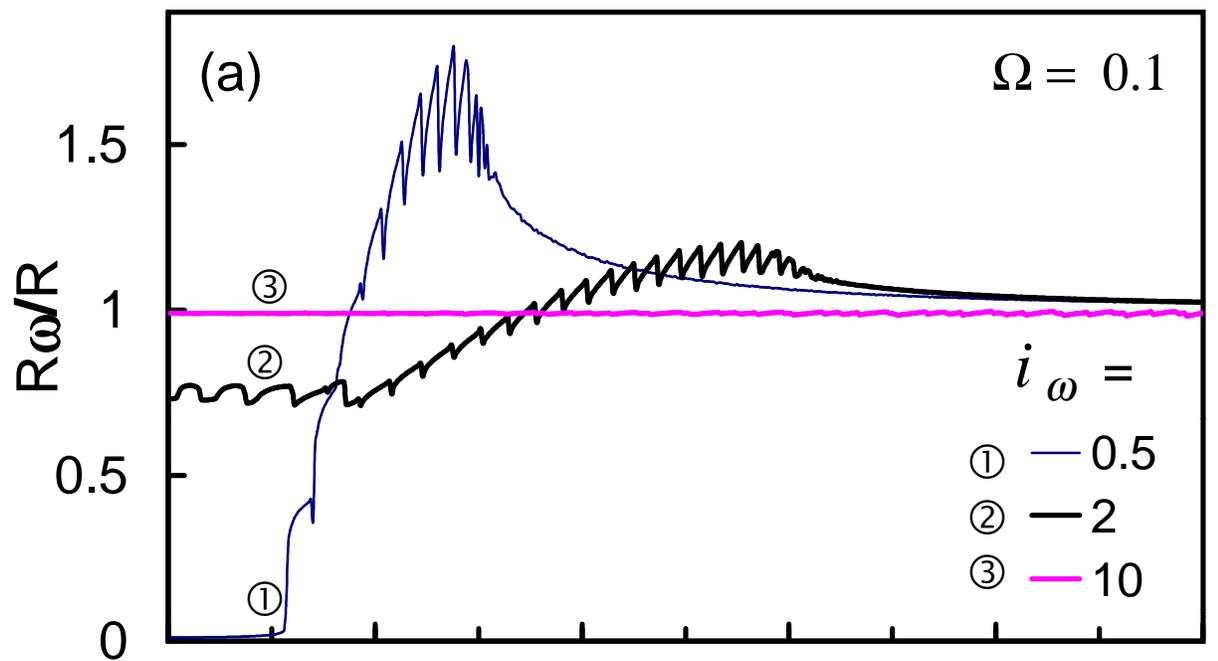
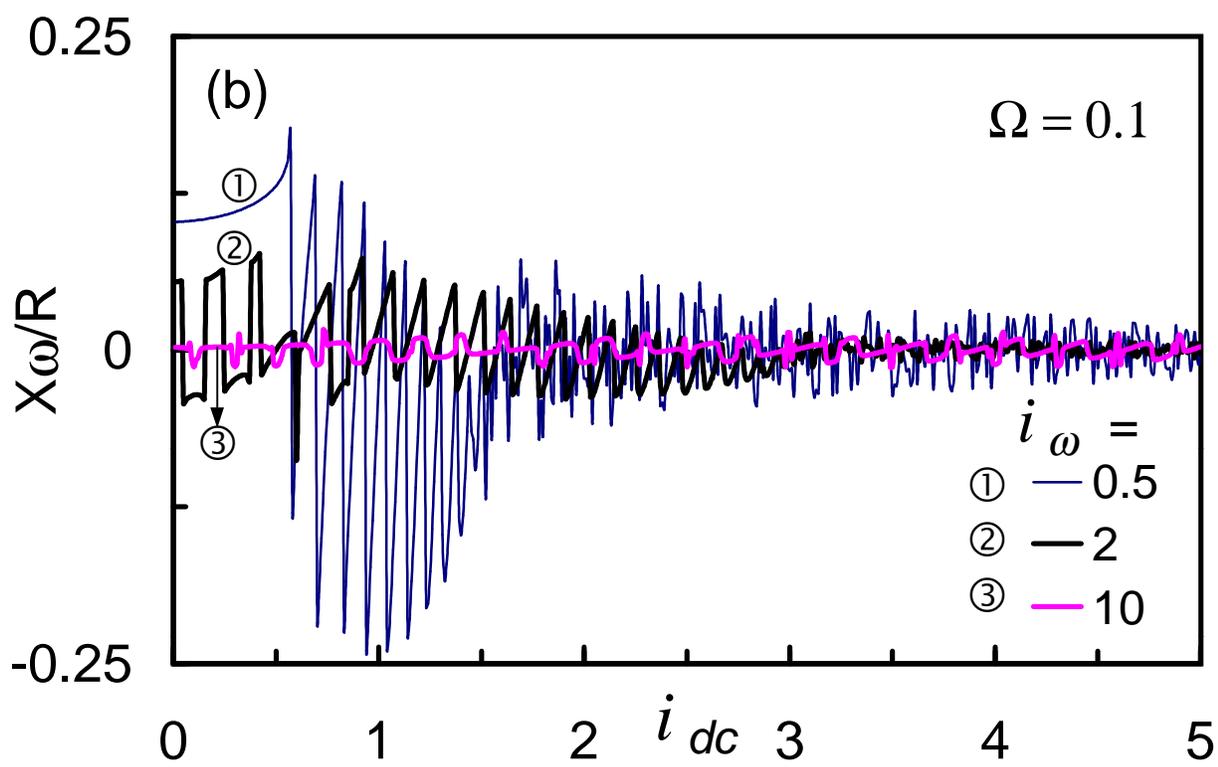
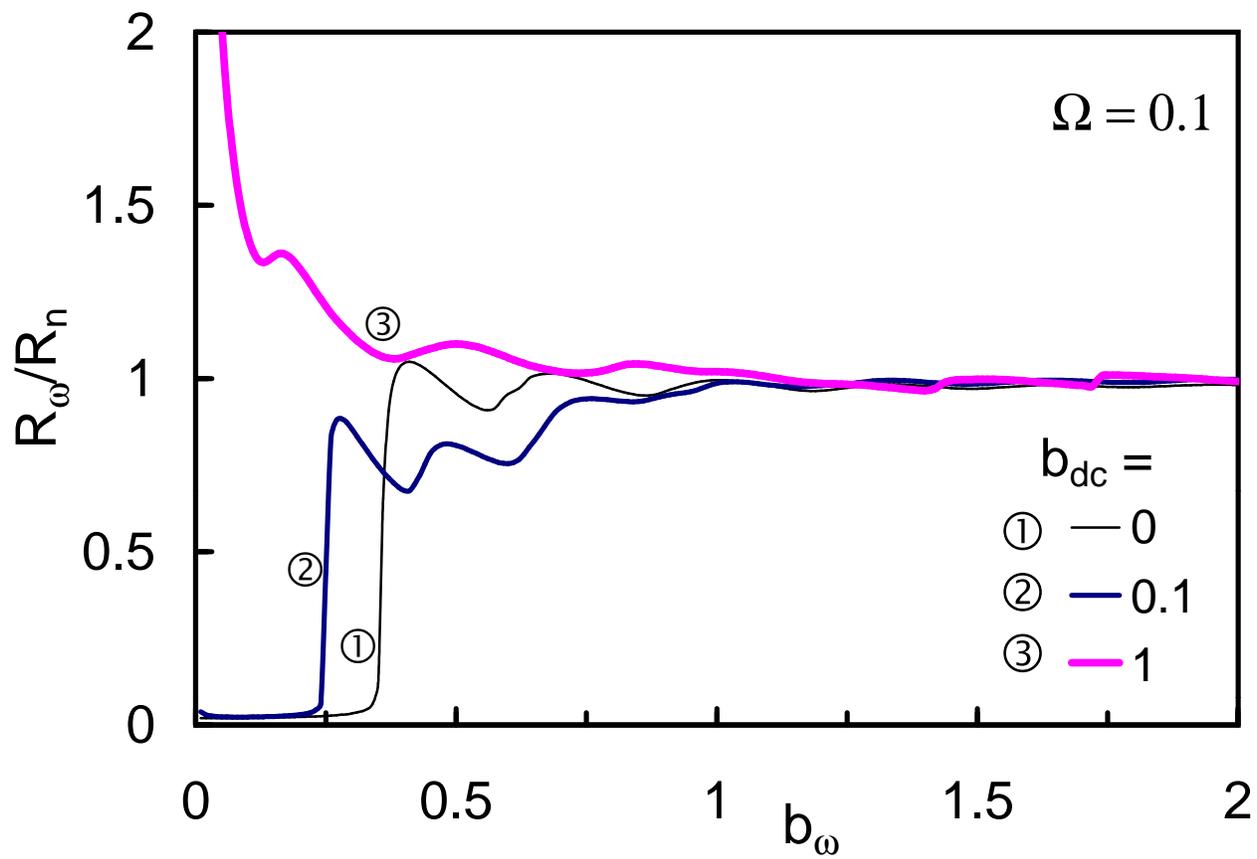
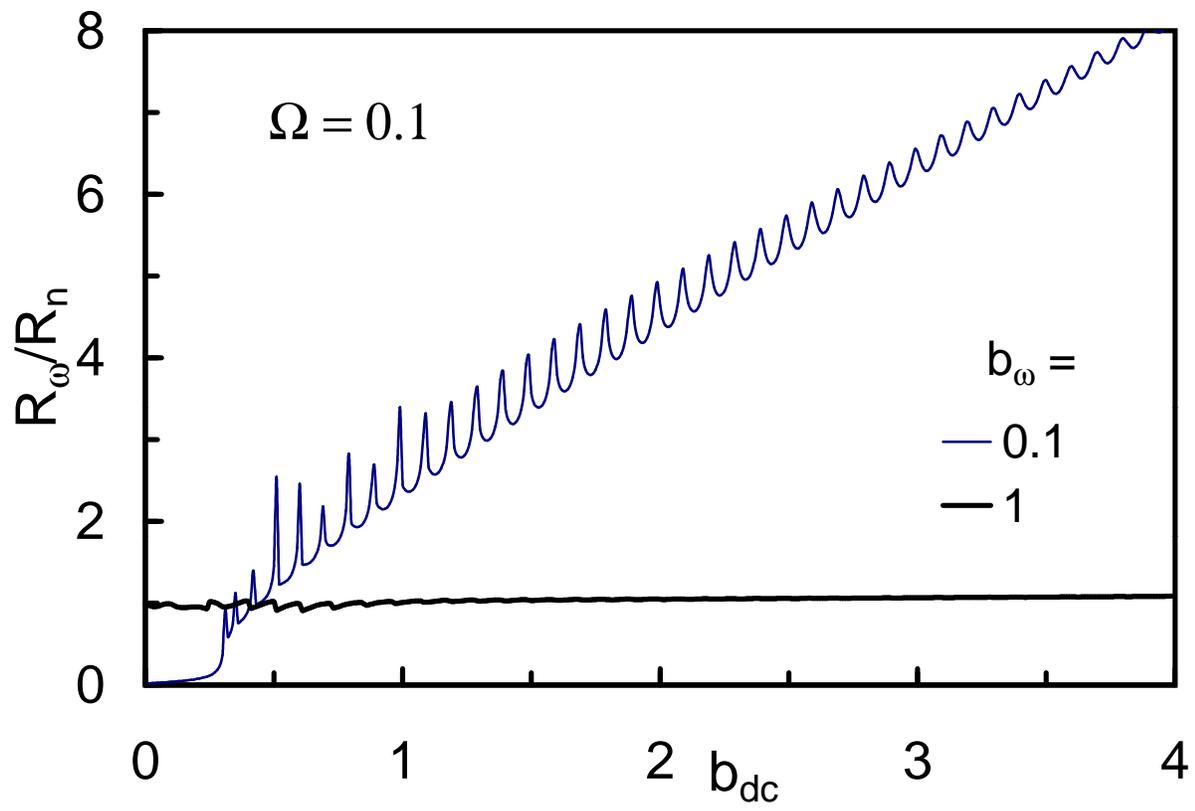
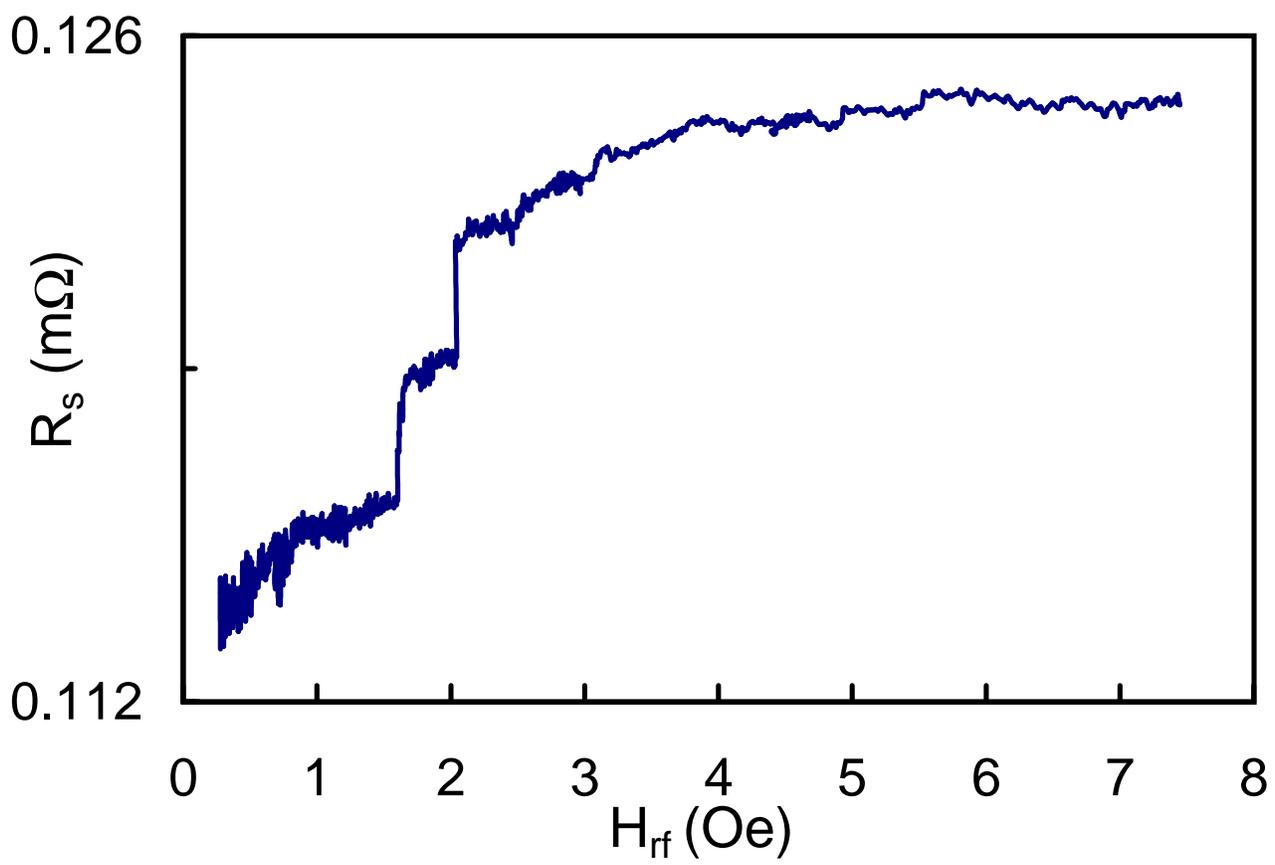
\end{document}